%% file: pfield.tex
\documentstyle[psfig2,psfrag,epsf,subfigure,makerobust,mncite]{mn}

\title
[Mapping potential fields on the surfaces of rapidly rotating stars]
{Mapping potential fields on the surfaces of rapidly rotating stars}
\author[G.A.J. Hussain, M. Jardine and A. Collier Cameron]
       {G.A.J. Hussain$^{1,2}$, M. Jardine$^1$ and A. Collier Cameron$^1$ \\
$^1$ School of Physics and Astronomy, University of St. Andrews,
 North Haugh, St Andrews KY16 9SS, UK \\
$^2$ Smithsonian Astrophysical Observatory MS-16, 
60 Garden Street, Cambridge MA 02138, USA \\
\hspace{1mm} ({\tt ghussain@cfa.harvard.edu}, {\tt mmj@st-andrews.ac.uk} and {\tt acc4@st-andrews.ac.uk})\\
}
\begin{document}
\input latex_macros
\maketitle
 
\begin{abstract}
 
We present a technique that combines Zeeman Doppler imaging (ZDI)
principles with a potential
field mapping prescription in order to gain more information about the
surface field topology of rapid rotators. 
This technique is an improvement on standard ZDI,
which can sometimes suffer from the suppression
of one vector component due to the effects of stellar inclination,
poor phase coverage or lack of flux from dark areas on the surface.
Defining a relationship beween the different vector components allows
information from one component to compensate for reduced information
in another.
We present simulations demonstrating the
capability of this technique and discuss its prospects.
 
\end{abstract}
 
 
\section{Introduction}
 
The solar-stellar analogy is commonly invoked to explain stellar activity
phenomena in terms of solar features such as prominences and flares
\cite{radick91}. A solar-type dynamo mechanism is also thought to operate
in other late-type stars.  However it is unclear if this exact same
mechanism will operate in stars covering a wide range of convection zone
depths, rotation rates and masses. Through the Mt Wilson H\&K survey,
which has monitored stellar rotation periods and activity cycles using
chromospheric emission variations spanning a period of over 30 years, we
know that many cool stars also display regular solar-type activity cycles.
However, other stars in the survey show irregular variations and some show
no cyclic patterns over the time span of the observations
 \cite{donahue96,baliunas95}.
\scite{saar99} and \scite{brandenburg98} find that the type
of dynamo activity appears to change with rotation rate and
age in a study using data from
the Mt Wilson survey as well as other photometric studies of cool stars.
 
Doppler imaging techniques can map $T_{\rm eff}$ flux 
distributions on the surfaces of rapidly rotating cool stars
\cite{cameron92doppler,piskunov90,rice89,vogt88}.
Starspot maps of cool stars taken a few rotation cycles apart allow
us to determine the differential rotation rate on rapid rotators
accurately \cite{donati97doppler,barnes2000pztel,petit2000HR1099,donati2000RXJ}.
As \scite{cameron00omega} report, these measurements lend support
to differential rotation models in which rapid rotators
are found to have a solar-type differential rotation pattern
\cite{kitchatinov99}.
Doppler maps of rapidly rotating stars show flux patterns 
which are very different to those seen on the Sun,
with polar and/or high-latitude structure often co-existing 
with low-latitude flux \cite{deluca97,strassmeier96table}.
This is hard to reconcile with the solar distribution 
where sunspots and therefore the greatest concentrations of
flux tend to be confined between $\pm 30^{\circ}$ latitude. 
Does this indicate different dynamo modes are excited in these stars ?
Models by \scite{granzer00} and \scite{schussler96buoy} explain 
the presence of mid-to-high latitude structure on young, rapidly rotating
late-type stars in terms of the combined effects of increased 
Coriolis forces and deeper convection zones working to ``pull'' 
the flux closer to the poles.
Even though \scite{granzer00} find that both  equatorial and polar flux 
can exist on T Tauri stars that have very deep 
convection zones, in terms of flux ``slipping'' over the pole, 
they find the emergence of low-latitude flux on K dwarf surfaces 
difficult to explain.

The technique of Zeeman Doppler imaging (ZDI) allows us to map
the magnetic field distributions on the surfaces of rapid rotators
using high resolution circularly polarized spectra \cite{semel89}.
Magnetic field maps of the subgiant component of the RS CVn binary,
HR1099 ($P_{rot}=2.8$d), and the  K0 dwarfs, AB Dor ($P_{rot}=0.5$d),
LQ Hya ($P_{rot}=1.6$d) have been presented in several papers
\cite{donati97doppler,donati99doppler,donati99lq}.
These maps all show patterns for which there is no solar counterpart:
strong radial and azimuthal flux covers all visible latitudes.
On AB Dor, strong unidirectional azimuthal field encircling the pole is
consistently recovered over a period of three years.
This is very different to what is seen on the Sun, where hardly any
azimuthal flux is observed at the surface and mean radial fields
are much smaller than those observed in ZD maps.
 
While ZDI is an important tool in measuring the
surface flux on stars, some questions about the
authenticity of the features in these Zeeman Doppler maps remain.
ZDI makes no assumptions about the field at the stellar surface;
the radial, azimuthal and meridional vectors are mapped
completely independently \cite{hussain99thesis,donati97recon,brown91zdi}.
For the true stellar field, however, the physics of the stellar interior
and surface will determine the relationship between the field components.
Additionally, poor phase coverage is found to lead to
an increased amount of cross-talk
between radial, meridional and azimuthal field components in ZDI maps
\cite{donati97recon}.
Due to the presence of starspots,
these maps are also flux-censored.
Indeed \scite{donati99doppler} and \scite{donati97doppler} find that the
low surface brightness regions on AB~Dor appear to correlate with the
strongest regions of radial magnetic flux.
For these reasons Zeeman Doppler (ZD) maps
may not present an accurate picture of the surface field on these stars.
 
By mapping potential fields on the surfaces of stars, we can evaluate
more physically realistic models of the surface flux distribution.
While we do not necessarily expect the field to be potential
locally, this assumption should be adequate on a global scale
\cite{demoulin93}.
\scite{jardine99pot} evaluate the consistency of Zeeman
Doppler maps for AB Dor with a potential field using maps 
obtained over three years.
Although they find that the potential field cannot reproduce
features such as the unidirectional azimuthal flux found
at high latitudes, 
they do find a good correlation between the ZD azimuthal map
and potential field configurations produced by extrapolating from 
the ZD radial map.
 
The technique presented in this paper reconstructs potential
field distributions directly from observed circularly polarized
profiles. This technique allows us to produce different configurations which
are more physically realistic and which provide more information
about the surface topology.

\section{The technique}
 
\subsection{Modelling potential fields}
 
The surface
magnetic field, {\bf B}, is defined such that {\bf B}=$- \nabla \Psi$.
The condition for a potential field,  $\nabla $x{\bf B}=0, is thus satisfied
and $\nabla $.{\bf B}=0 reduces to Laplace's equation:
\begin{equation}
\nabla \Psi^{2} =0.
\end{equation}
 
At the surface, the radial field, $B_{r}$, azimuthal field, $B_{\phi}$,
and meridional field, $B_{\theta}$, can be expressed in terms of
spherical harmonics:
 
\begin{equation}
B_{r} = - \sum^{N}_{l=1} \sum^{l}_{m=-l}[la_{lm} - (l+1)b_{lm}]P_{lm}(\theta) e^ {im\phi}
\label{eq:magrad}
\end{equation}

\begin{equation}
B_{\phi} = - \sum^{N}_{l=1} \sum ^{l}_{m=-l} [a_{lm} + b_{lm}]\frac{P_{lm}(\theta)}{\sin\theta} ime^{im\phi}
\label{eq:magaz}
\end{equation}
 
\begin{equation}
B_{\theta} = - \sum^{N}_{l=1} \sum ^{l}_{m=-l} [a_{lm} + b_{lm}]\frac{d}{d\theta}P_{lm}(\theta) e^{im\phi}.
\label{eq:magmer}
\end{equation}

Here $P_{lm}(\theta)$ is the associated Legendre function at each latitude,
$\theta$.

We map the components, $a_{lm}$ and $b_{lm}$. 
They are allowed to vary until they produce 
surface field vector distributions, $B_{r}$, $B_{\phi}$ and $B_{\theta}$,
that fit the observed polarized profiles.
The procedure by which the circularly polarized dataset is predicted
using a $B_{r}$, $B_{\phi}$ and $B_{\theta}$ image is described in
Section~\ref{sec:pol}.
As with conventional Doppler imaging, this problem
is ill-posed and many different $a_{lm}$ and $b_{lm}$ image configurations
can fit an observed dataset to the required level of $\chi^{2}$.
A regularizing function is thus required to reach a unique solution and 
we use maximum entropy in this capacity \cite{gull84,skilling84}.
The maximum entropy method has been used in several Doppler imaging
codes and enables us to produce a unique image
which has the minimum amount of information that is required to fit the
observed dataset within a certain level of misfit
(as measured by the $\chi^{2}$ statistic) 
\cite{cameron92ccp7,piskunov90,rice89}.

\subsection{Maximum entropy}
\label{sec:maxent}

\begin{figure}
\psfig{file=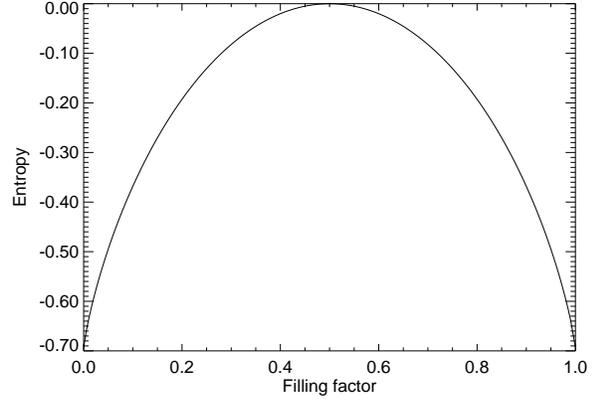,width=3.0in}
\caption{The power in each image pixel, $i$, (in each $lm$ mode) 
is allowed to vary between $-B_{max}$ (i.e. filling factor, $f_i=0.0$) 
and $+B_{max}$ ($f_i=1.0$), where $B_{max}=1000$G.
The entropy is at a maximum value for a zero magnetic field ($f_i=0.5$).
In the absence of information the code will push to the 
maximum possible value of entropy. Hence it weights equally 
for both positive and negative polarities.}
\label{fig:zdi_maxent}
\end{figure}

We store magnetic field strength in terms of a filling factor model.
The power in each mode is defined as a 
fraction of a maximum value of 1000~G. 
This is taken to be the upper limit for which the weak field 
approximation can still be applied (see Section~\ref{sec:pol}).
Magnetic flux values between the lower and upper limits, 
-1000~G and 1000~G, are mapped onto
flux filling factors, $f_i$, between 0.0 to 1.0 
(hence $f_i=0.5\equiv0.0$~G).

\begin{figure*}
\psfig{file=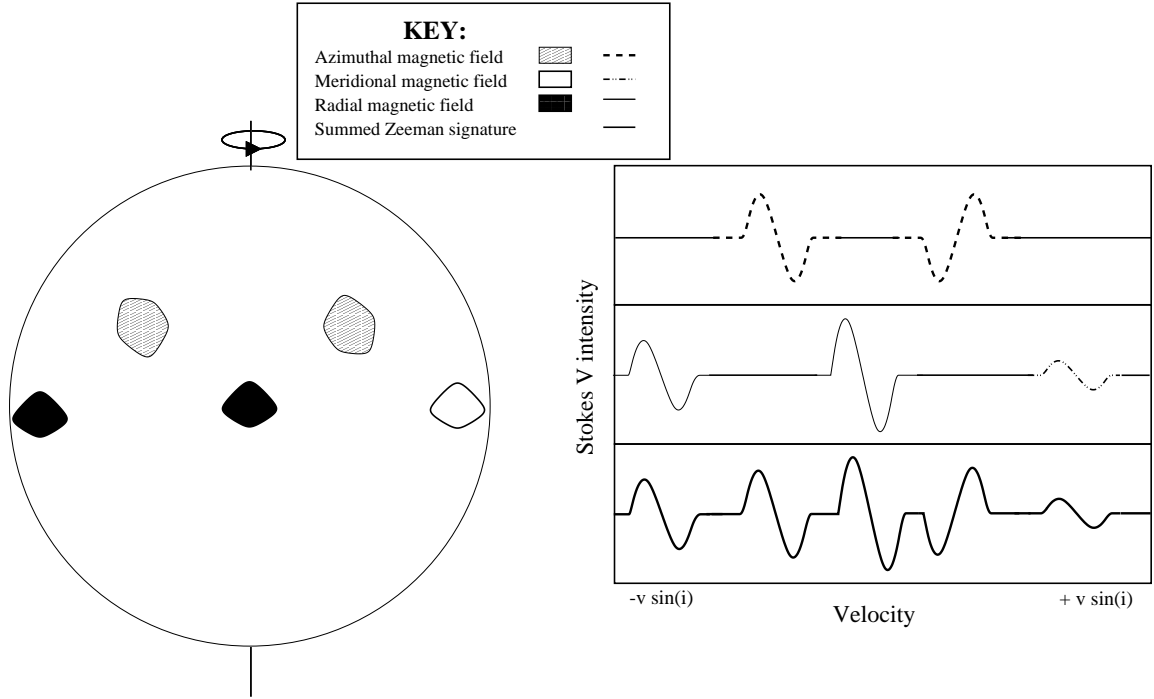,width=6.0in}
\caption{A schematic diagram illustrating how circularly polarized
spectral signatures from magnetic field spots vary depending on
the field vector. The contributions from radial, azimuthal and
meridional spots to the summed Stokes V spectral signatures are represented by 
different line types (as shown in the key). }
\label{fig:zdischem}
\end{figure*}

For each mode, $i$, the entropy, $S(f)$, 
is then defined in terms of flux filling factor, $f_i$, and 
image weights, $w_i$:

\begin{equation}
S(f) = - \Sigma w_{i}[f_{i}log \frac{f_{i}}{d} + (1-f_{i})\log\frac{1-f_{i}}{1-d}].
\label{eq:zdient}
\end{equation}
The default image value, $d=0.5$ (i.e. 0~G).
Fig.~\ref{fig:zdi_maxent} shows how this form of entropy pulls equally in both
directions to zero field images.

On assigning equal weights to all $lm$ modes, 
we find that the reconstructed images 
are preferentially pushed to the highest, most complex distributions.
However we want to obtain
the simplest possible potential field image which fits 
the observed spectral dataset.
In order to obtain this reconstruction,
we assign weights according to 
the number of nodes at the surface for each mode (i.e. each $lm$ combination):

\begin{equation}
\frac{1}{W(l,m)}= [(2|m| + 1)(l-|m|+1)]^{n}.
\label{eq:weights}
\end{equation}

The choice of $n$ determines how strongly the image
is pulled to the minimum number of nodes:
$n=1$ means that the weights are proportional to  
the number of nodes on the surface; $n=2$ represents
the surface area density of nodes. 

It should be noted that this weighting scheme simplifies the
amount of information we reconstruct about the stellar surface.
The choice of $n$ in Eqn.~\ref{eq:weights} does not affect the simple
reconstructions 
but is found to produce more accurate reconstructions in the case of
the more complex topologies. This is discussed in more detail in the next
section. 
Using the solar-stellar analogy, it is likely that the actual 
distribution of magnetic regions is much more complex. The
weighting scheme used here allows us to test the code and obtain a better idea
about fundamental limitations inherent in the scheme. 
When dealing with real stellar spectra, 
a more realistic weighting scheme may be to push the reconstructed image
towards a power-law distribution. 
The detailed effect of different weighting schemes on real stellar spectra 
will be the subject of a later paper.

\subsection{Modelling circular polarization}
\label{sec:pol}

The procedure used to calculate predicted polarized line profiles is 
described more fully in \scite{hussain00comps}.
Briefly, the model used is that of \scite{donati97recon}. 
This involves making the following assumptions about the 
intrinsic line profile:
\begin{itemize}
\item that it is constant over the surface; 
\item it can be modelled using a Gaussian; 
\item the weak field approximation is valid.
\end{itemize}

Within the weak field regime,
when Zeeman broadening is not the dominant line broadening mechanism,
the Stokes V profile contribution, $V_i$,
at a wavelength, $\lambda$,  from a magnetic region, $s$,
is found to behave as follows:
\begin{equation}
 V_s(v)\propto g \lambda \frac{\partial I_{s}(v)}{\partial v}.
\label{eq:weakfield}
\end{equation}
Here $g$ is the average effective Land\'e factor over all the
considered transitions;
$I_s$ is the unpolarized intensity line profile contribution from region, $s$;
and $v$ is the velocity shift from the rest wavelength.
The weak field approximation is found to be valid for mean 
(least squares deconvolved) profiles 
up to magnetic field strengths of 1 kG \cite{bray65}.

Through rotational broadening, flux contributions from different
regions on the stellar surface are separated in velocity space.
As with conventional Doppler imaging, the velocity excursions
of the line profiles indicate the region of the stellar surface 
where the signature originates.
High-latitude field signatures are confined to the centre of the 
line profile and distortions at the line profile wings are caused by 
magnetic fields in the low latitude regions.
Circularly polarized (Stokes V) profiles are only sensitive to the 
line-of-sight component of the magnetic field. 
Fig.~\ref{fig:zdischem} is a schematic diagram showing how magnetic
field spots with similar parameters but different vector orientations
have differing Stokes V contributions depending on their position
on the projected stellar disk.
The Stokes V signature
from an azimuthal magnetic field spot switches sign as the spot crosses the
centre of the projected stellar disk. Stokes V signatures from 
radial field spots do not show any switch, 
just scaling in amplitude with limb darkening. 
Similarly, low-latitude meridional field spot signatures do not switch sign.
However, as the line-of-sight projection of low-latitude meridional
vectors is suppressed, the circularly polarized signature is also reduced.
Hence amplitude variations in a time-series of circularly polarized
spectra indicate both the strength of the magnetic flux at the surface
as well as the vector orientation of the field.

For an exposure at phase $k$,
the Stokes V specific intensity contribution, $V_s$, for a surface
element pixel, $s$, to a data point in velocity bin, $j$,
is determined as shown below for each field orientation:
\begin{equation}
D_{jk} = \Sigma \frac{B_{s}}{B_{max}}[V(\Delta\lambda_{sjk},\mu_{sk}) \cos\theta_{sk} ].
\label{eq:forward}
\end{equation}
Here $B_{s}$ is the magnetic field value in pixel,  $s$,
at a particular field orientation
and $B_{\mbox{max}}=1000$~G. 
$V(\Delta\lambda_{sjk})$ represents the Stokes V
specific intensity contribution for a 1000~G field pixel 
that has been Doppler-shifted 
by the instantaneous line-of-sight velocity of the pixel.
The term, $\cos\theta_{sk}$, is the projection into the line-of-sight 
of image pixel, $s$, at phase, $k$.

\section{Simulations and reconstructions}
\label{sec:simulatns}
We use a similar approach to \scite{donati97recon} and demonstrate
the capabilities of this technique
by comparing potential field and Zeeman Doppler 
reconstructions for several different input datasets.

All reconstructions presented here have been fitted to a reduced $\chi^2=1.0$. 
For both sets of images, we used stellar parameters based 
on the active K0 dwarf, AB Dor: 
$P_{rot}=0.51479$~d; inclination angle, $i=60^{\circ}$; 
\vsini$=90$\kmsec; \vrad$=0$\kmsec.

\subsection{Simple images}
The line and image parameters used for this test reconstruction
are listed below.\\
Image: $a_{1,1,}$~=~300.0~G; $b_{1,1}$~=~-300.0~G; $b_{3,2}$~=~250~G
Line: $\lambda_{\circ}$~=~5000~\AA, S/N~=~1.0E+05, spectral resolution~=~3~\kmsec \\

\begin{figure*}
\centerline{\mbox{\psfig{file=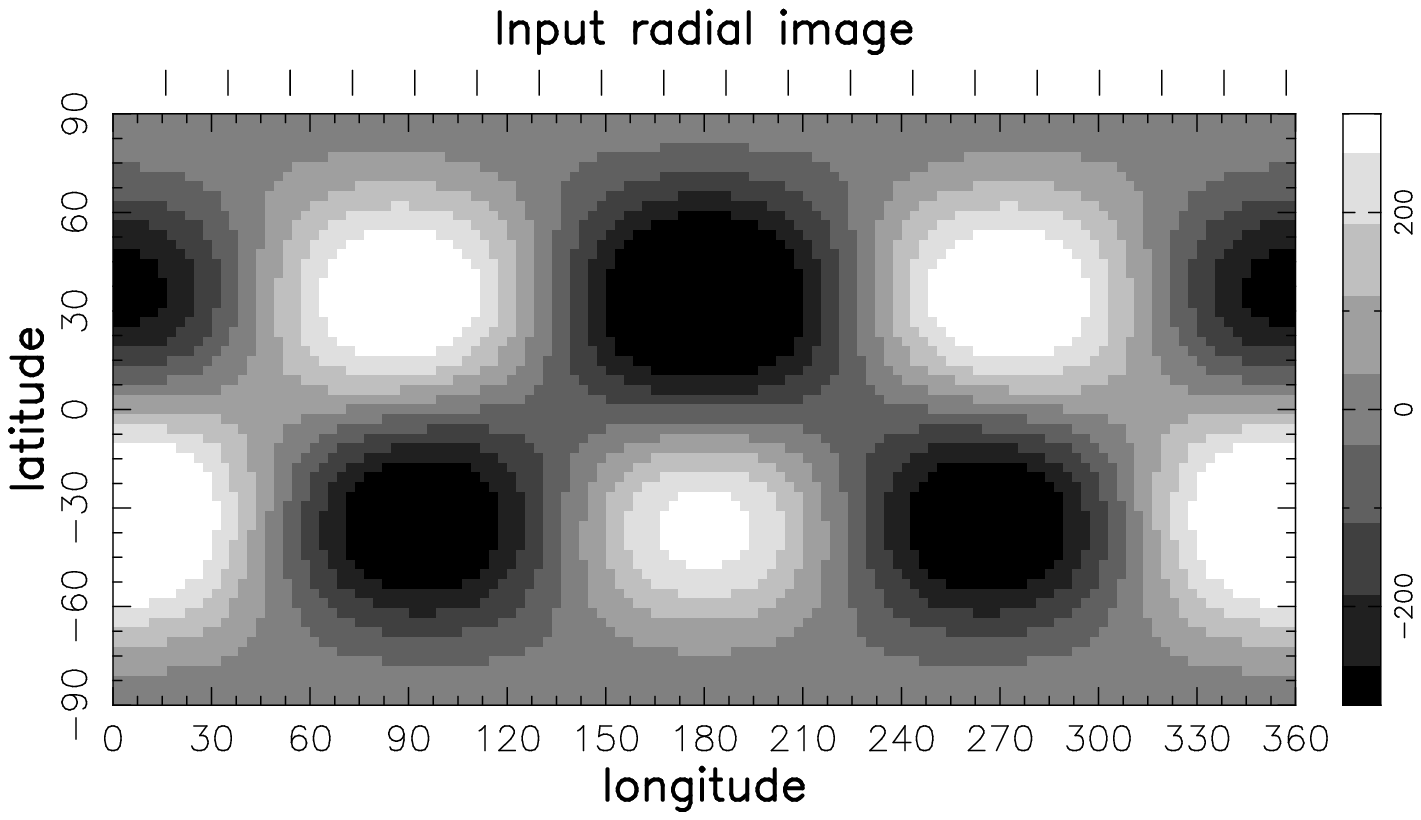,width=5.5cm} 
\psfig{file=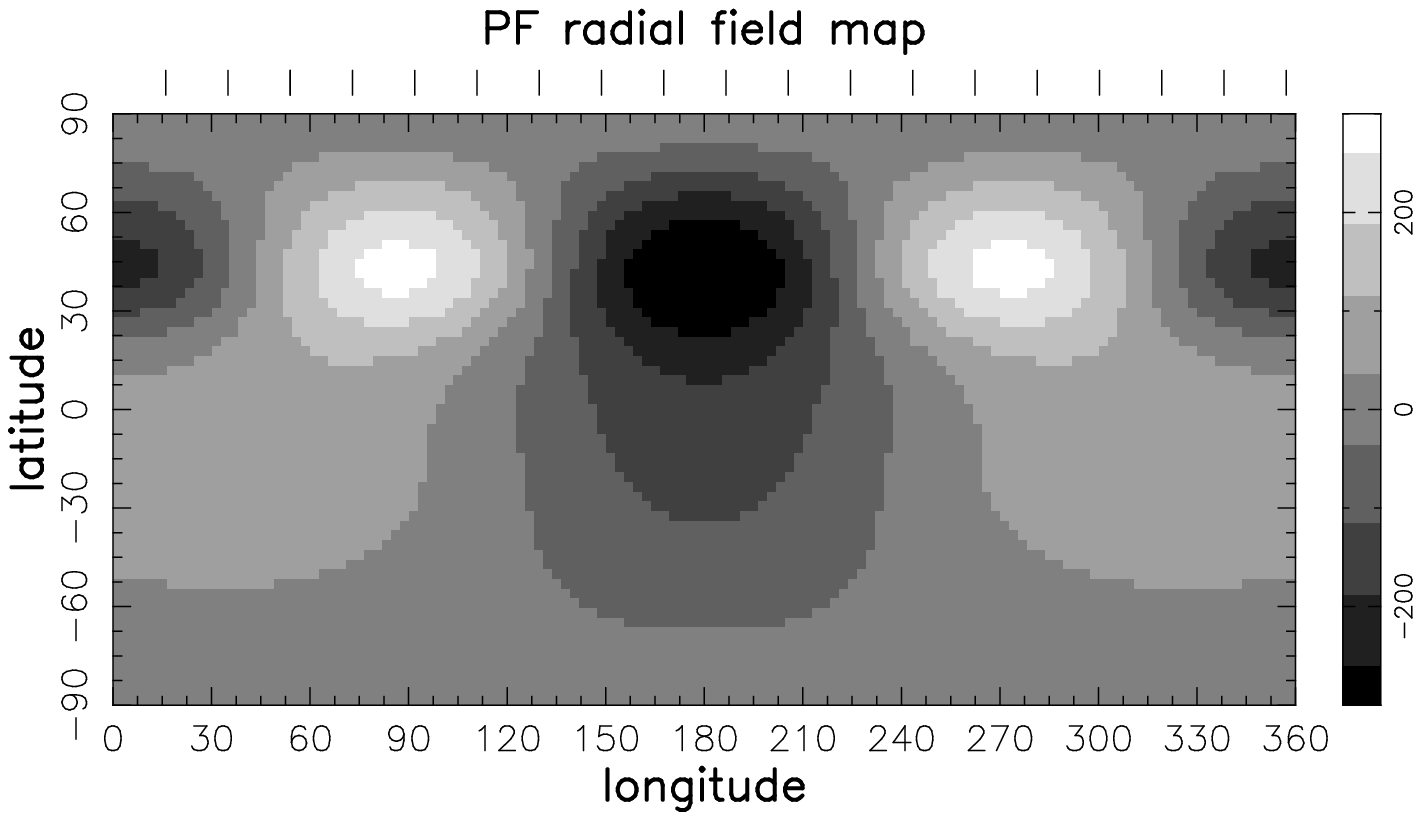,width=5.5cm}
\psfig{file=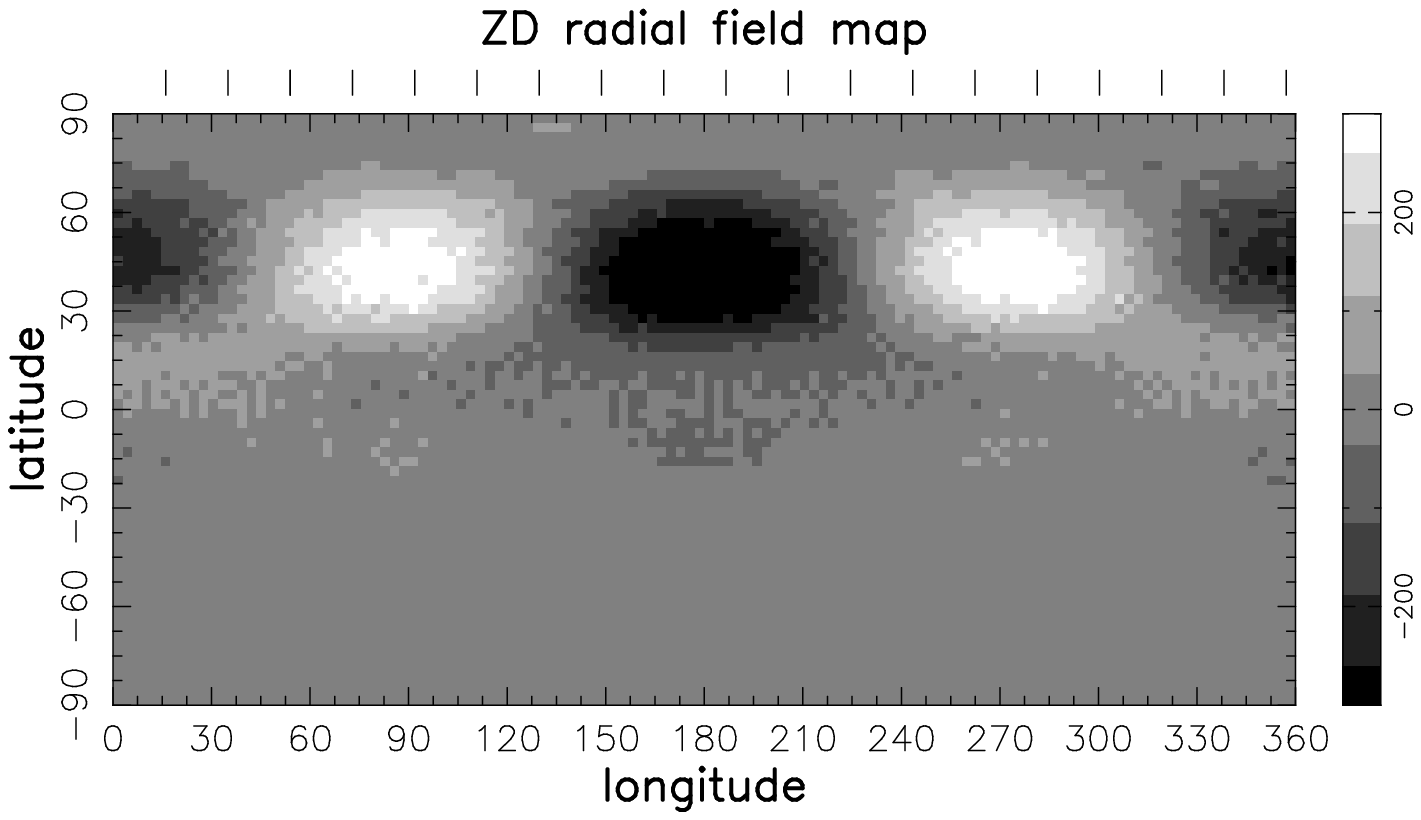,width=5.5cm}}}
\centerline{\mbox{ \psfig{file=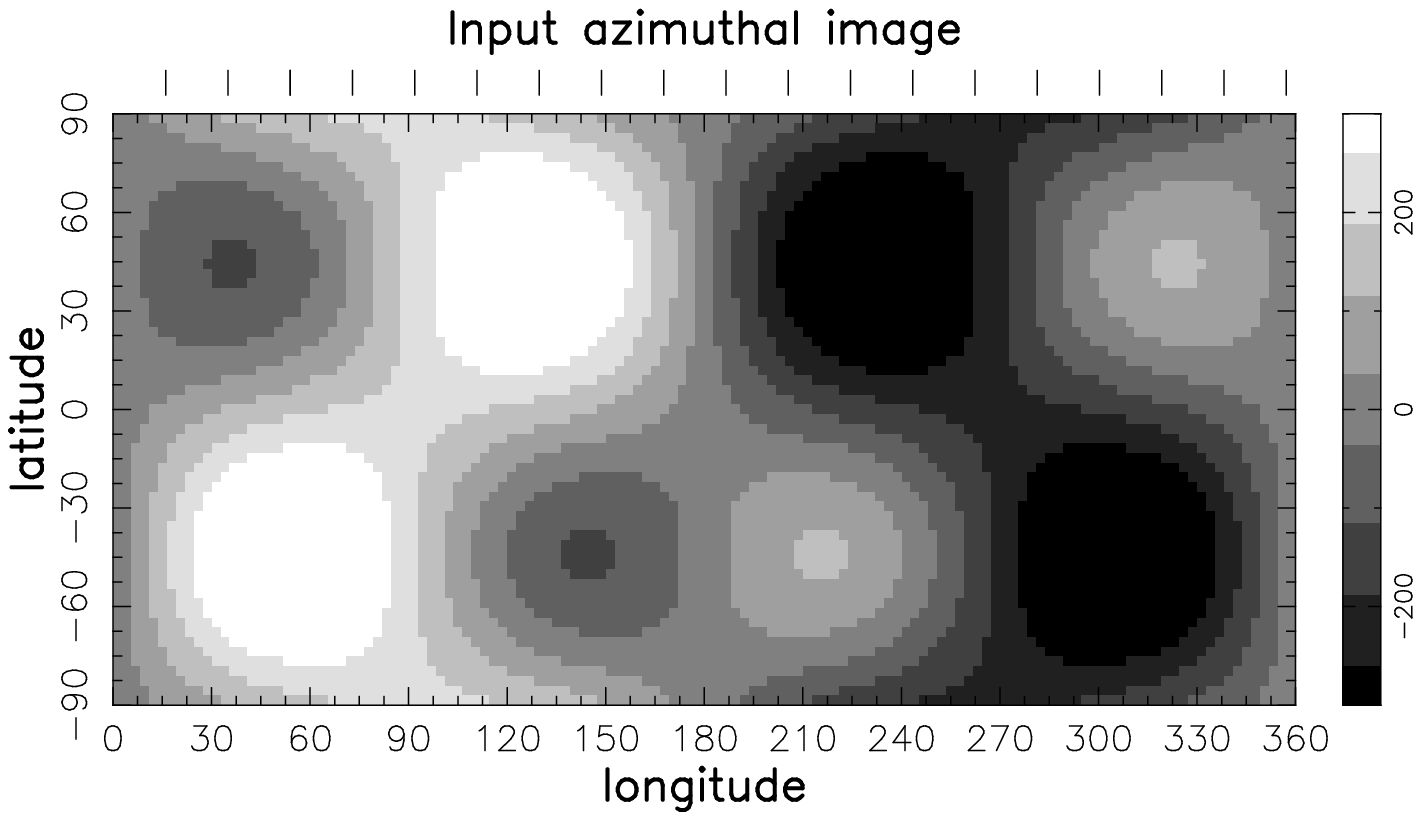,width=5.5cm}
\psfig{file=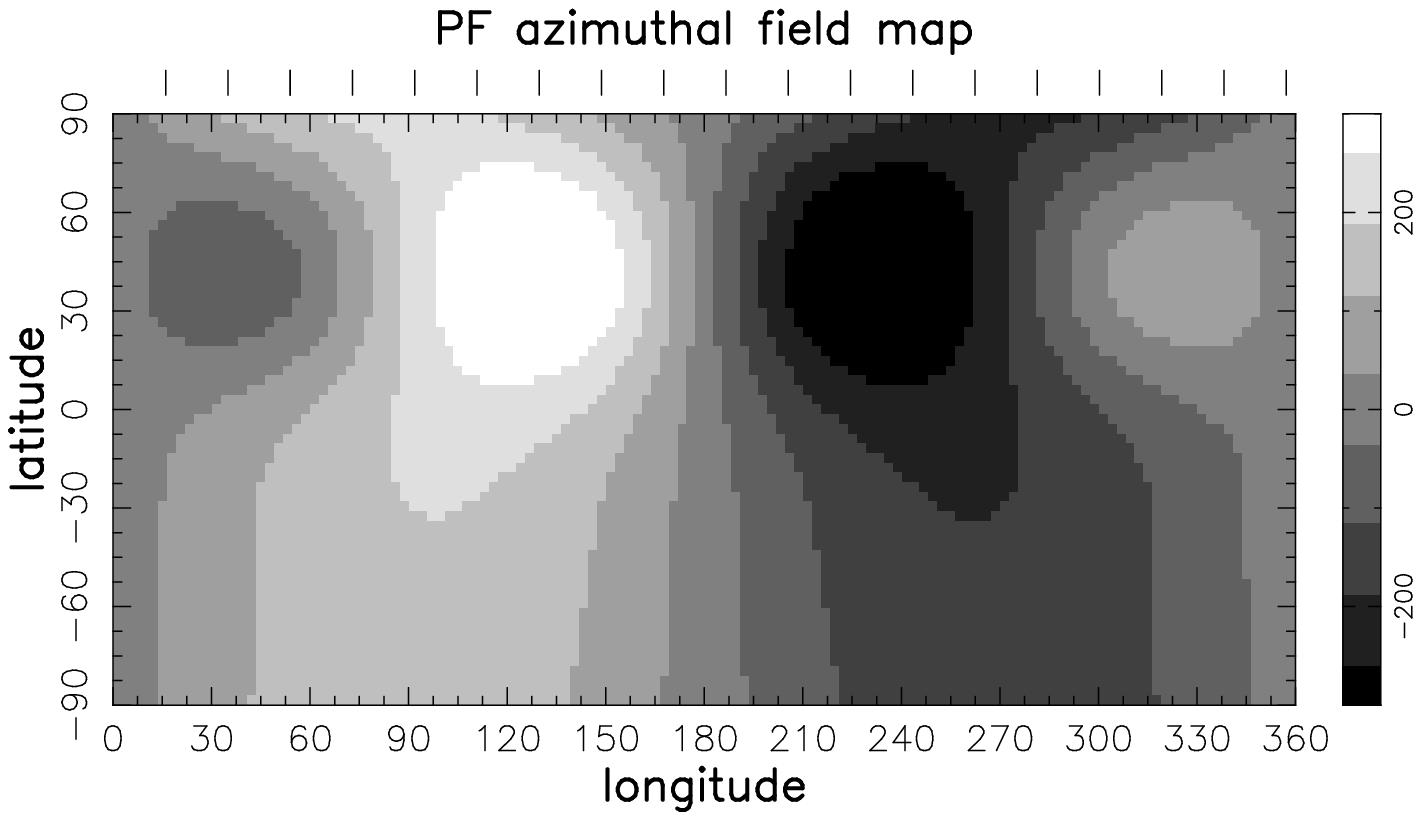,width=5.5cm}
\psfig{file=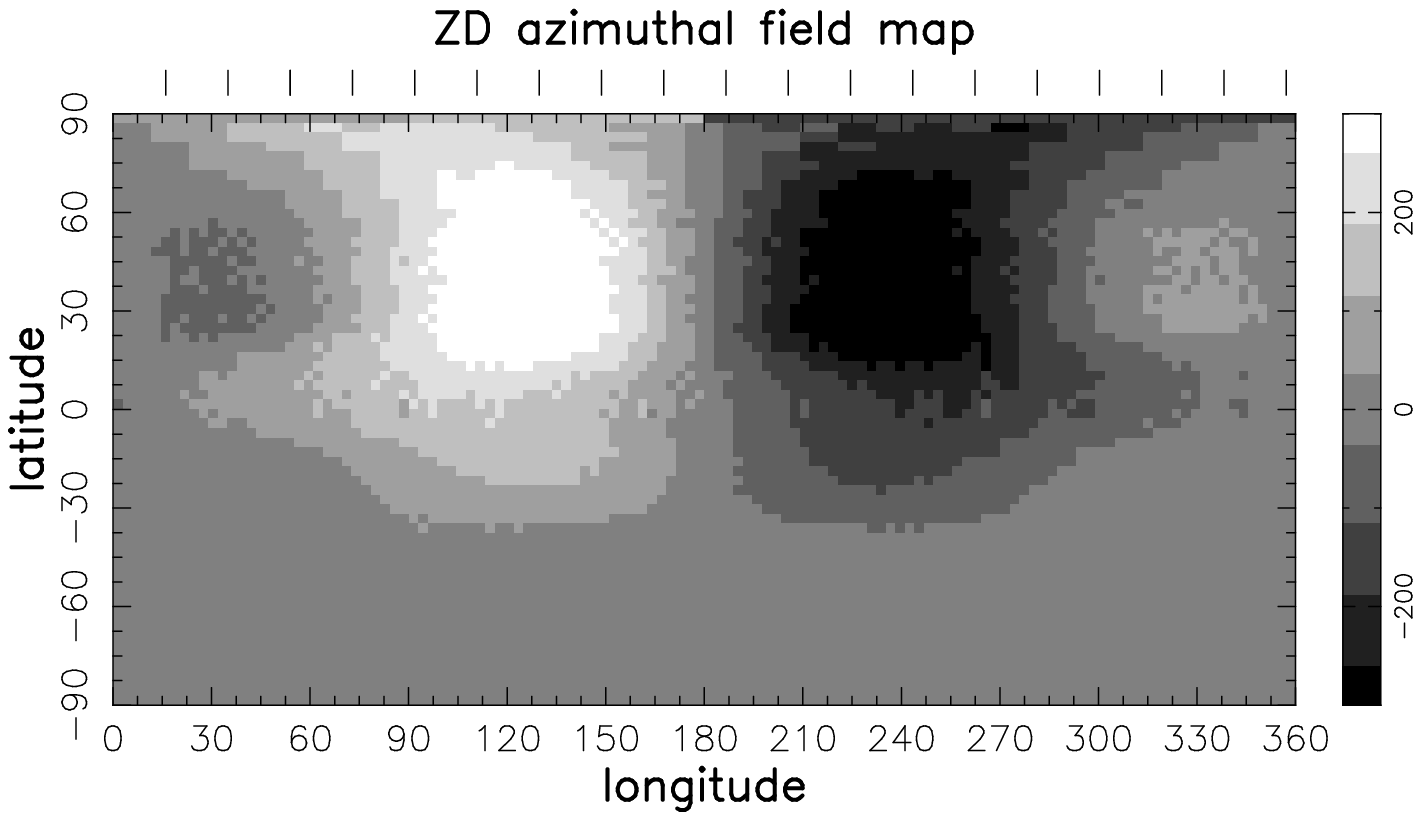,width=5.5cm}}}
\centerline{\mbox{ \psfig{file=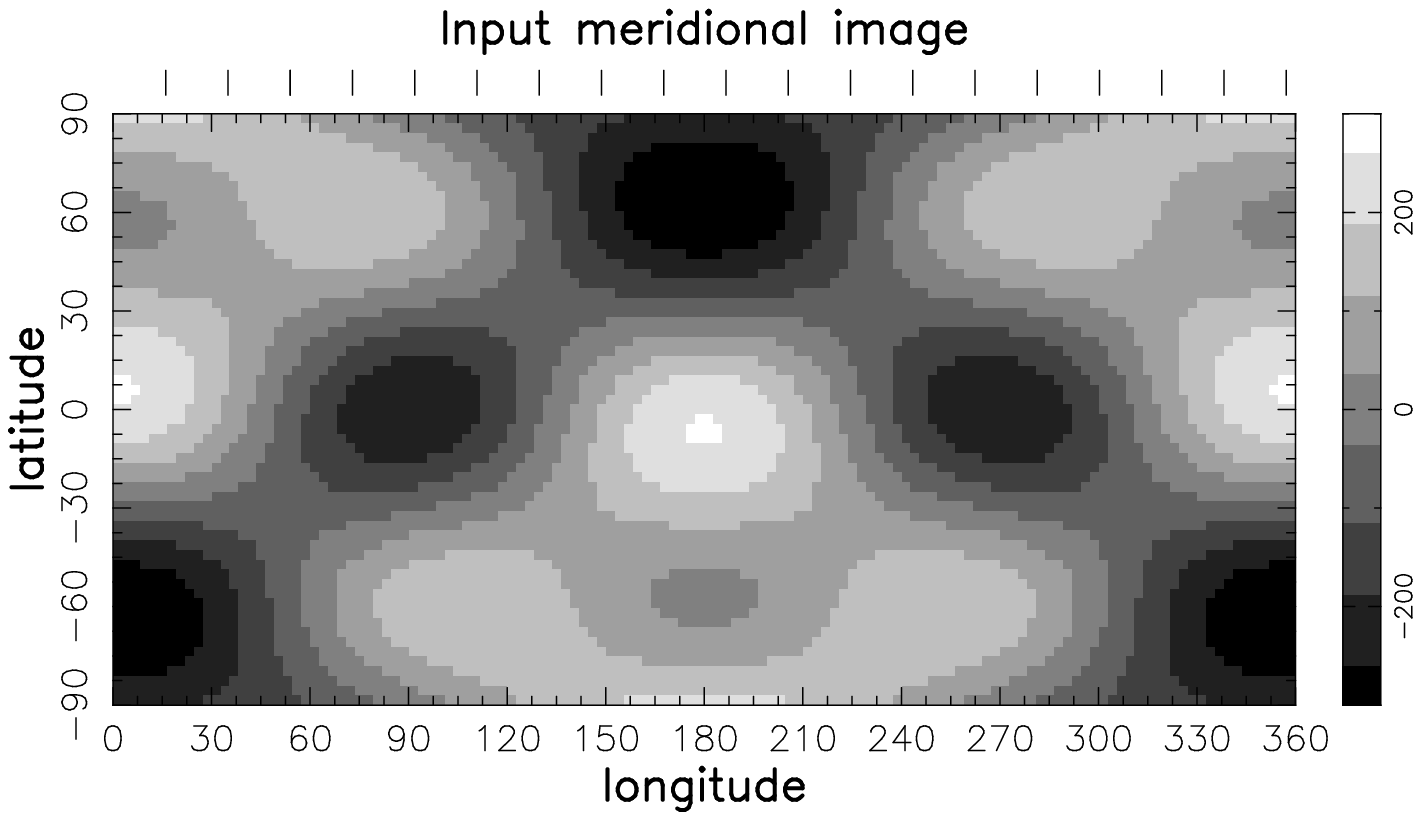,width=5.5cm}
\psfig{file=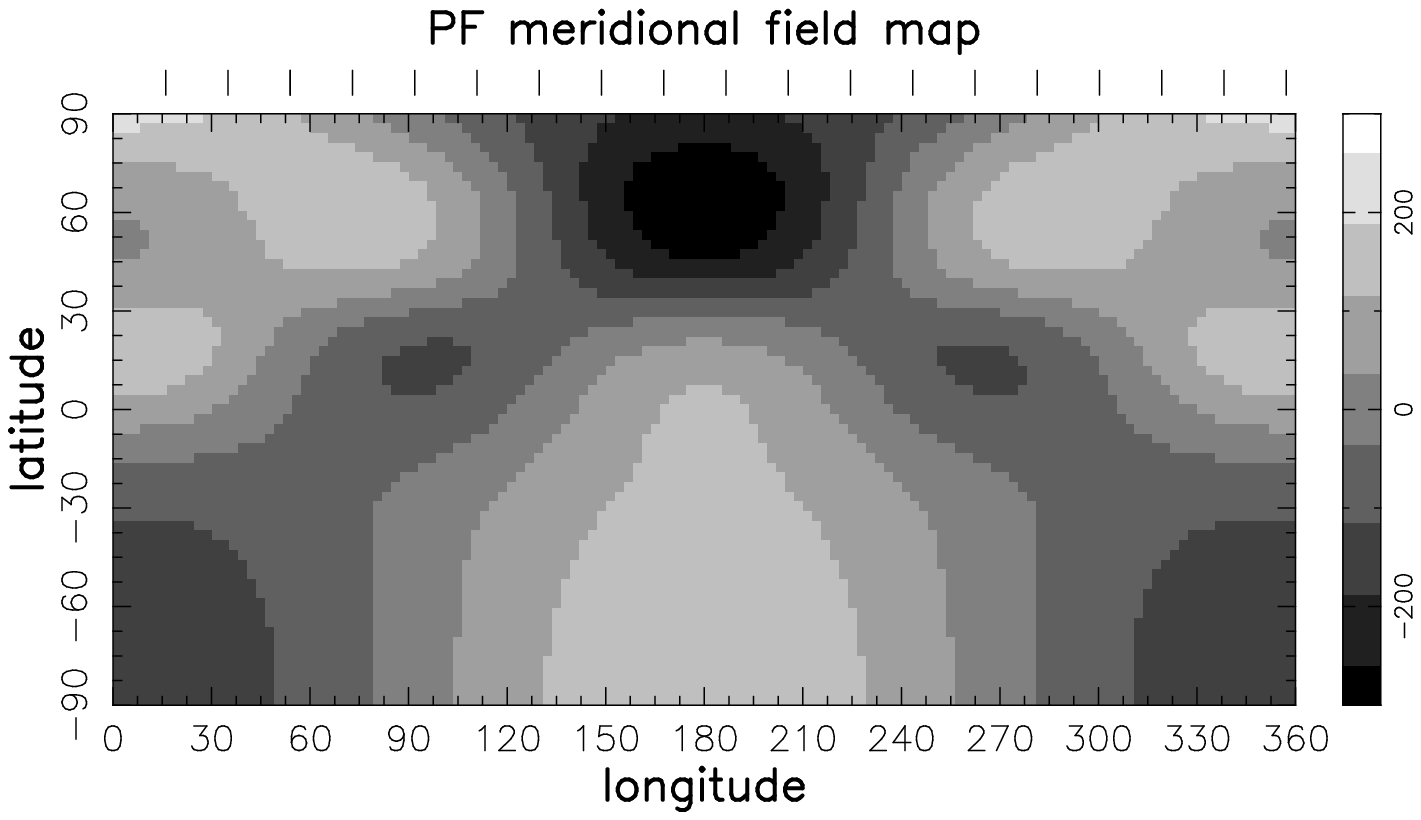,width=5.5cm}
\psfig{file=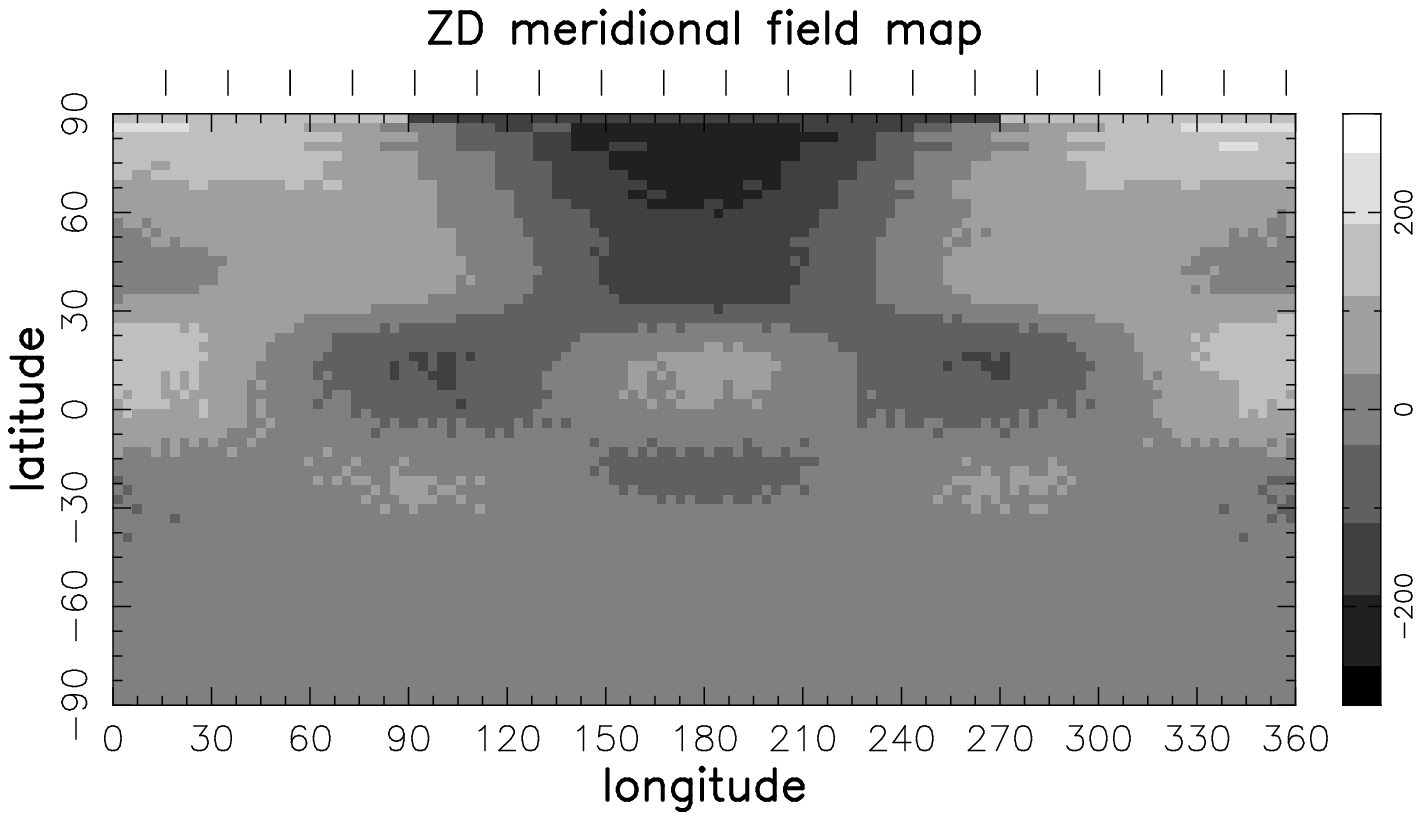,width=5.5cm}}}
\caption{Input surface radial, azimuthal and meridional field maps
calculated using $a_{1,1}=300.0$ G; $b_{1,1}=-300.0$ G; $b_{3,2}=250$ G.
Greyscales represent the magnetic flux strength and
the tick marks denote the observation phases.
Potential surface radial, azimuthal and meridional 
field maps are plotted in the middle column. 
Zeeman Doppler maps are plotted in the right column.
Both sets of potential field maps and Zeeman Doppler maps fit the 
model spectral dataset to reduced $\chi^2=1.0$}.
\label{fig:simple}
\end{figure*}

\begin{figure}
\centerline{\psfig{file=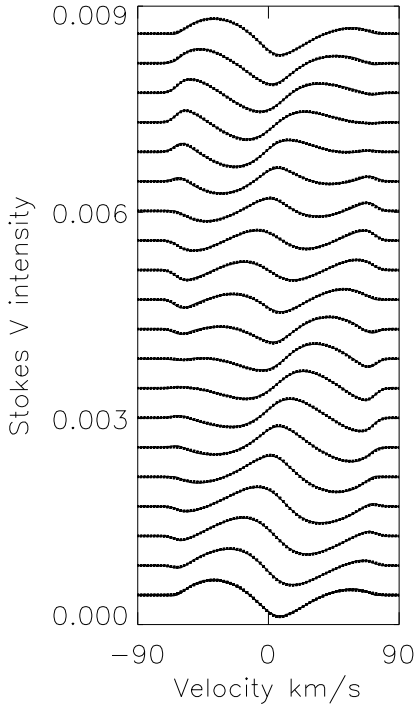,width=5.4cm}}
\caption{Model Stokes V spectra generated using the input image shown
in Fig.~3 are represented by the error bars. 
Spectra are plotted in order of increasing phase in steps of 0.05 phase.
The predicted profiles from the potential field reconstructions are
shown by the solid lines.  The Zeeman Doppler predicted line profiles
are not plotted here but fit the model spectra to a similar accuracy.}
\label{fig:simplefits}
\end{figure}

The ZD maps were produced using the ZDI code described in 
\scite{hussain99thesis}.
The input surface field, reconstructed potential field and Zeeman Doppler
maps  are shown in  Fig.~\ref{fig:simple}.
The model Stokes V spectra and fits from the potential
field reconstructions are plotted in Fig.~\ref{fig:simplefits}.
The field in the visible hemisphere ($-30^{\circ}<\theta<90^{\circ}$) 
is considerably smeared out in both the potential field and 
ZD reconstructions.
There is also clearly some cross-talk between the high-latitude 
meridional and azimuthal components in the ZD maps. 
This is because circularly polarized signatures
from both high-latitude azimuthal and meridional fields
change sign over a full rotation phase. This leads to 
ambiguity in the reconstructions.
Low-latitude radial and meridional fields suffer from a similar ambiguity
as the line-of-sight vector component from these regions does not change 
sign thus leading to cross-talk.
As the potential field maps show, the additional constraint
of a potential field leads to less ambiguity in the reconstructions.
While the polarities are reconstructed correctly for the potential 
meridional field map, the flux strengths are greatly suppressed.
This is because the meridional field vector is almost completely 
suppressed at low latitudes \cite{hussain99thesis,donati97recon} and 
hence the information needed to reconstruct them is largely absent in
the circularly polarized profiles

The agreement between input and reconstructed images has been evaluated 
by calculating the reduced $\chi^{2}$ value between each input and
reconstructed map. The flux values for each magnetic field map were
first normalised according to pixel area, such that the sum of
all pixel areas is equal to $4 \pi$. Where $N$= total number of pixels
in each map, $I_i$= corresponding pixel in the input map and
$R_i$=reconstructed map pixel;  the reduced $\chi^{2}$ value 
is then evaluated is defined as $\chi^{2}/(N-1)$ where:
\[ \chi^{2} = \sum_{i=1}^{N} (I_{i} - R_{i})^{2}. \]

\begin{table}
\caption{Reduced chi-squared values comparing input and reconstructed images.
The first column lists which images they correspond to,  
the remaining three columns indicate the level of
agreement between reconstructed images for each field orientation. 
Clearly the potential field reconstructions show  better agreement with the
input images compared to the Zeeman Doppler reconstructions.}

\vspace{0.5cm}
\begin{tabular}{llll}
\hline
Image  & $\chi^{2}_{rad}$ & $\chi^{2}_{az}$ & $\chi^{2}_{mer}$ \\
Full Inp - ZD  & 40.6  & 57.2 & 15.5 \\
Full Inp - PF  & 36.1  & 12.4 & 6.6 \\
50\% Inp - ZD  & 50.3  & 54.2 & 17.2 \\
50\% Inp - PF  & 39.5  & 13.8 & 9.3 \\
\hline
\end{tabular}
\protect\label{tab:chisq1}
\end{table}

As shown in Table~\ref{tab:chisq1} potential field reconstructions
in all three field vectors show  a better level of agreement
 with the input maps compared to the ZD maps. The reduction
in cross-talk between surface field vectors is reflected by the
much lower reduced $\chi^2$ values in the 
PF meridional and azimuthal field maps.

\subsubsection{Bad phase coverage}

\begin{figure*}
\centerline{\mbox{ 
\psfig{file=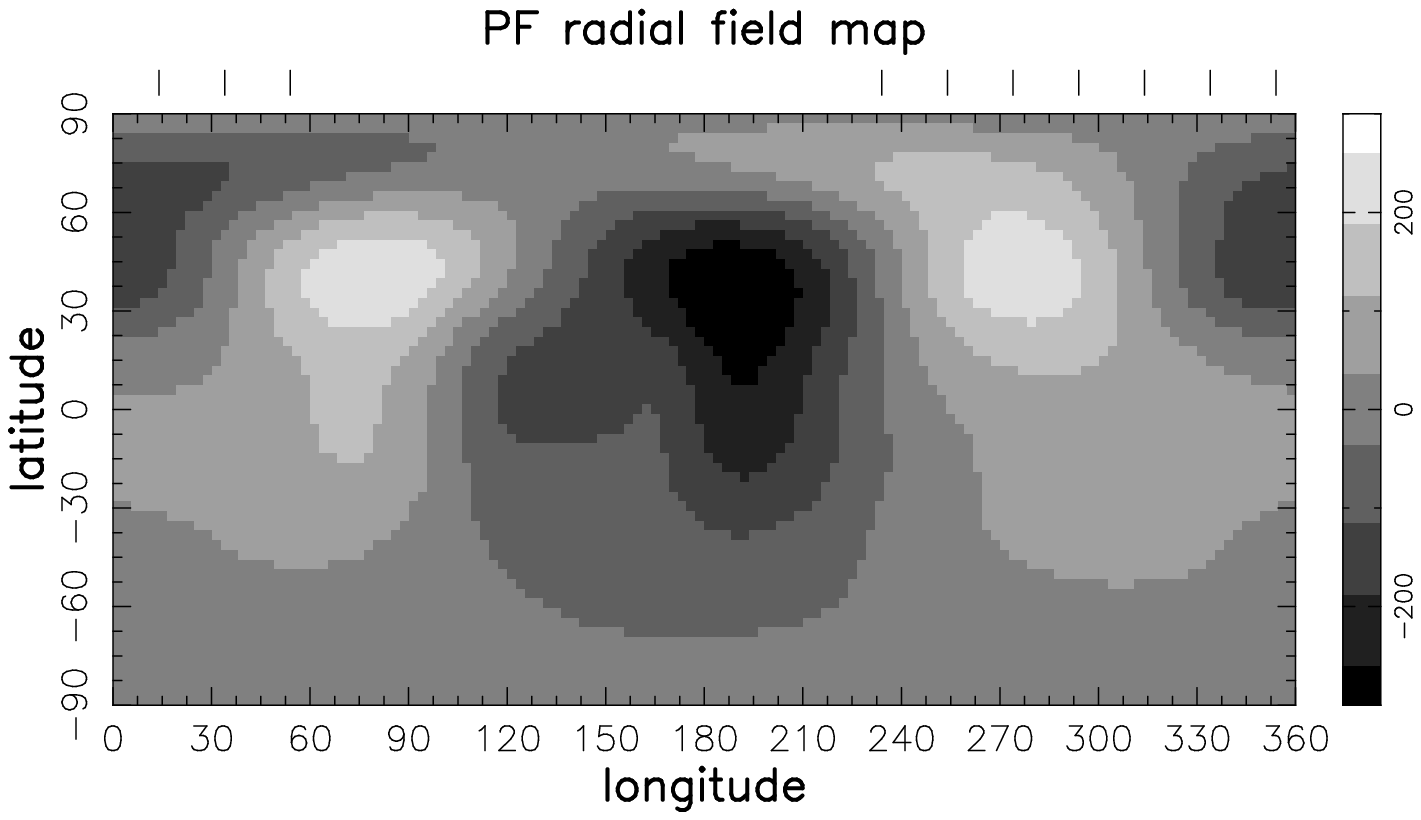,width=5.5cm}
\psfig{file=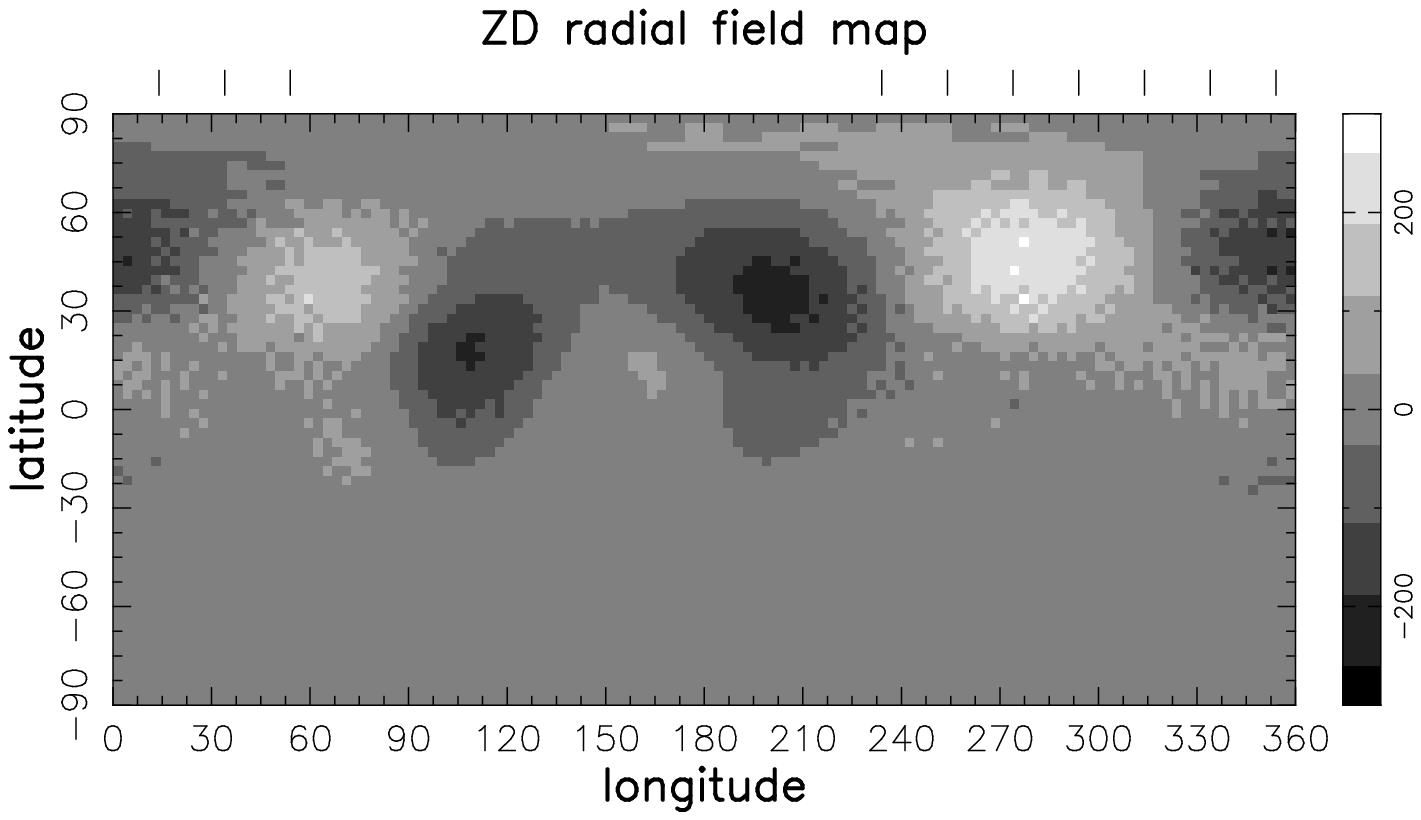,width=5.5cm}}}
\centerline{\mbox{ \psfig{file=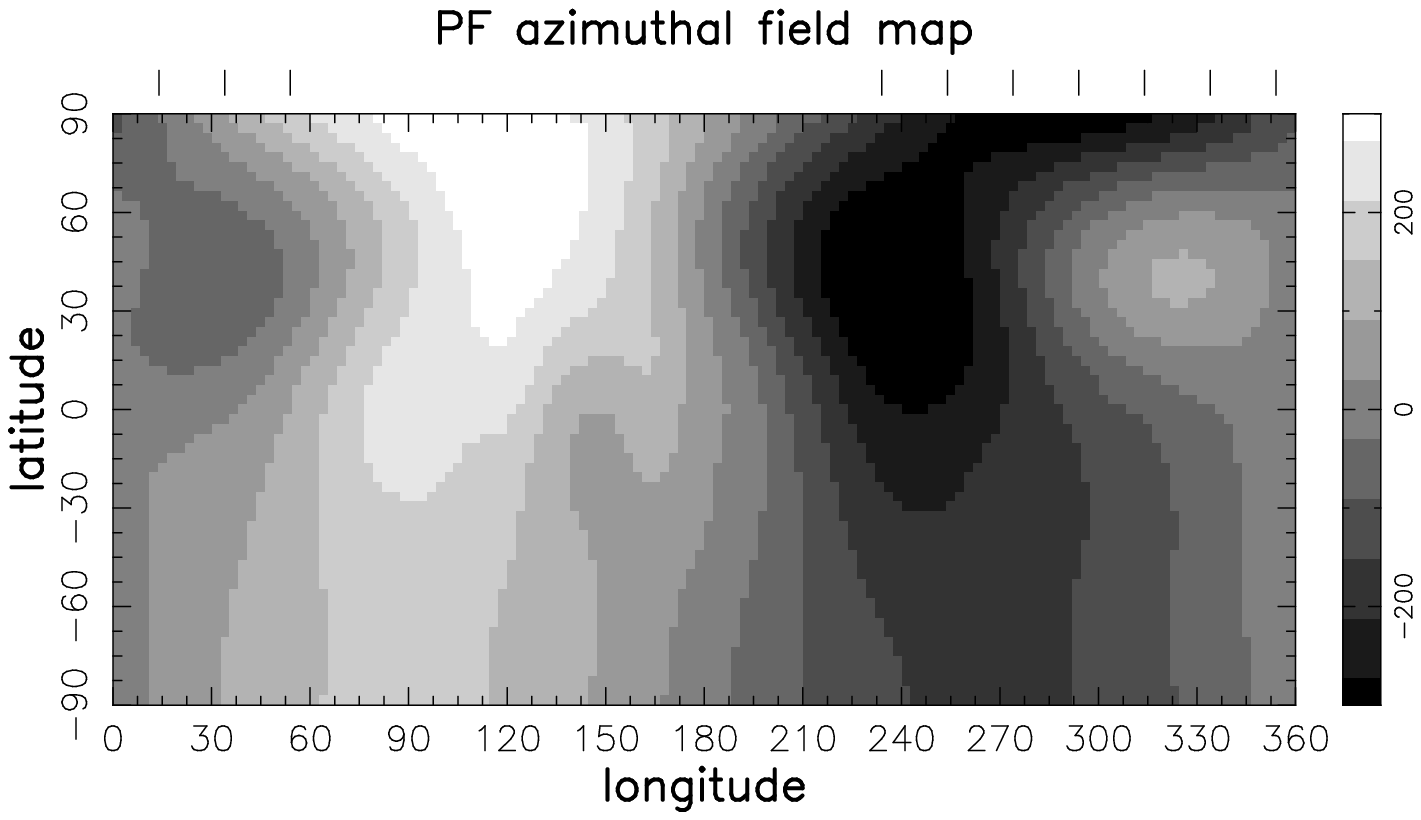,width=5.5cm}
\psfig{file=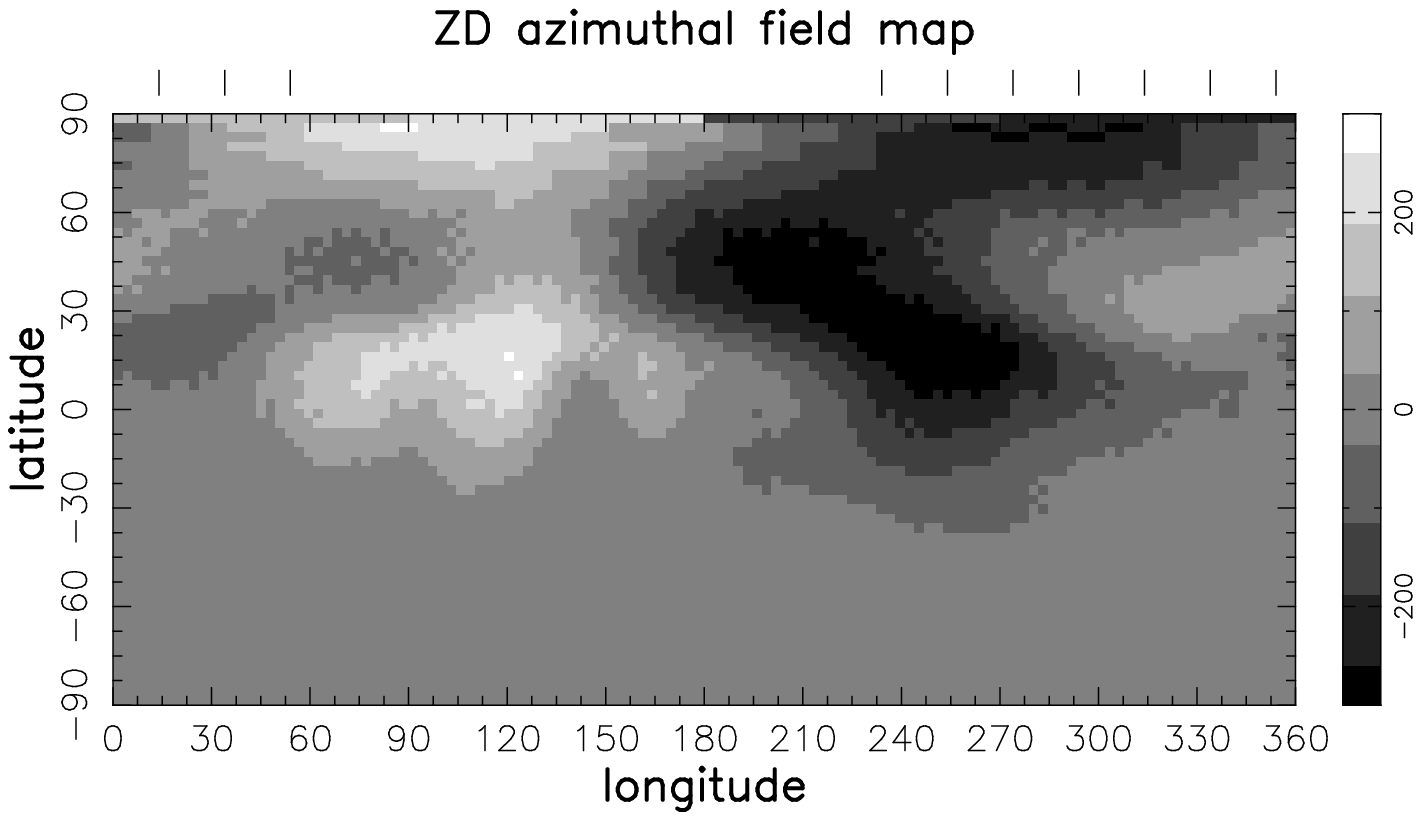,width=5.5cm}}}
\centerline{\mbox{
\psfig{file=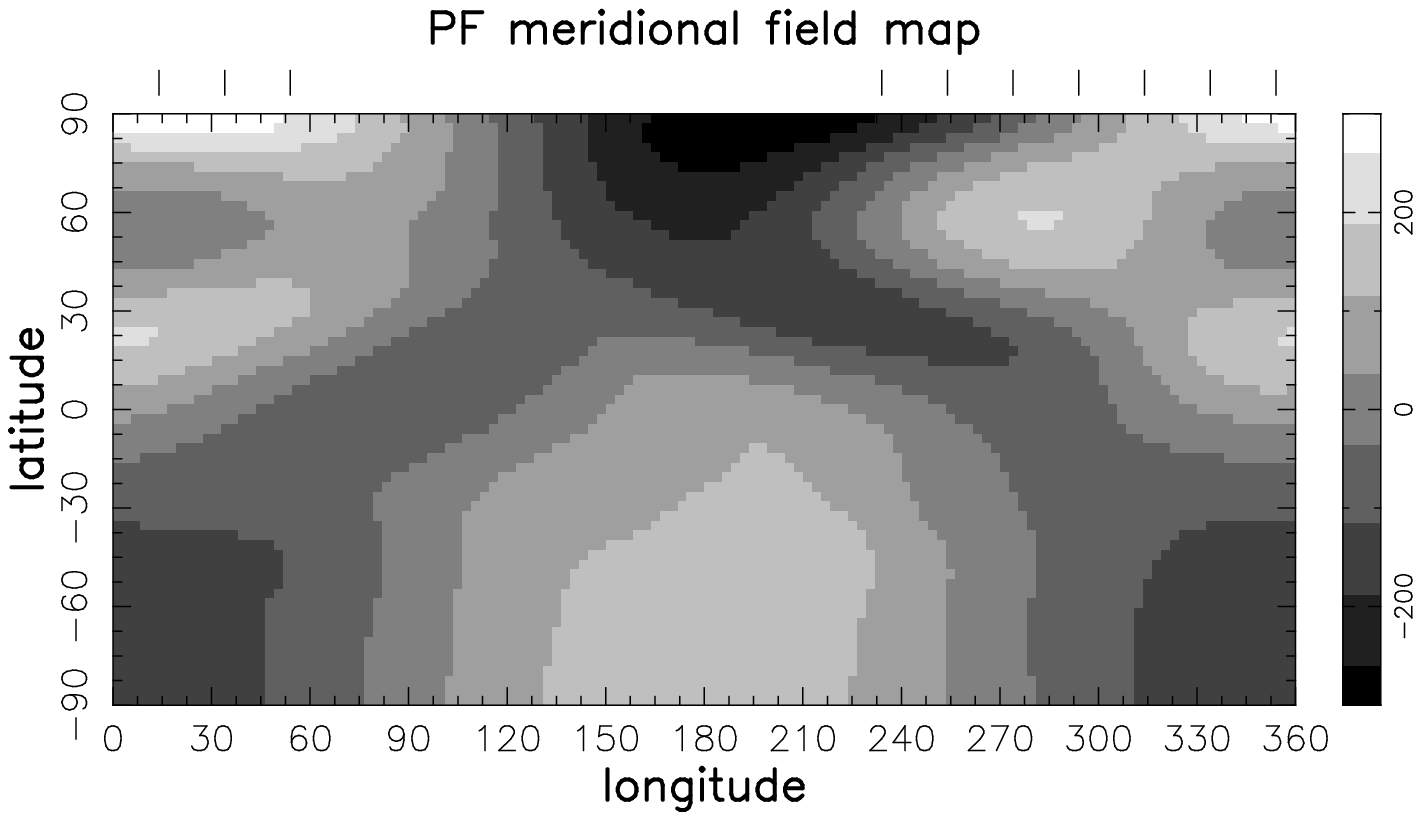,width=5.5cm}
\psfig{file=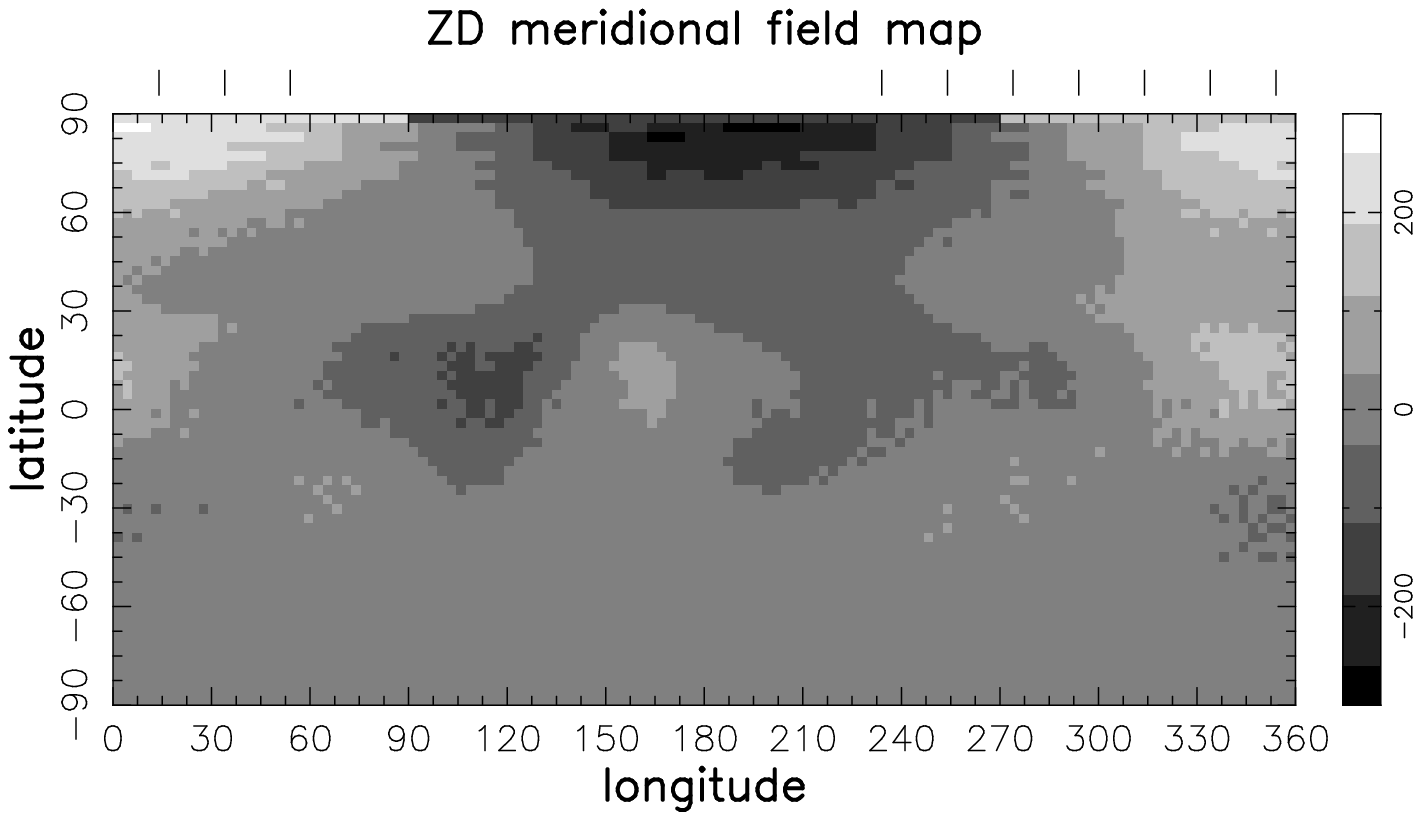,width=5.5cm}}}
\caption{
Maps showing the effects of poor phase coverage.
Potential radial, azimuthal and meridional field
maps are plotted in the left column and 
ZD maps are plotted in the right column.
As before, all maps fit the spectral dataset to a 
reduced $\chi^2=1.0$. Observation phases are
represented using tick marks.}
\label{fig:simplephase}
\end{figure*}

In the case of high inclination stars, 
bad phase coverage leads to an increased amount of 
ambiguity in ZD maps, especially  
between low-latitude radial and meridional field vectors \cite{donati97recon}. 
We test the effects of bad phase coverage on potential field reconstructions
and compare them with ZD maps for the same image and line parameters 
as those listed above but for
a dataset with only 50\% phase coverage.
The potential field and ZD maps for this poorly sampled dataset
are shown in Fig.\ref{fig:simplephase}. 
As before, all reconstructions fit the observed dataset 
to a reduced $\chi^{2}=1.0$.

We would expect the amount of cross-talk between field vectors to be 
much reduced in the potential field maps compared with ZD maps.
The reason for this is that the line-of-sight component 
is strongest approximately $45^{\circ}$ away from the 
centre of the stellar disk
Hence, when the radial field contribution is reduced 40$^{\circ}$-50$^{\circ}$ 
away from the disk centre due to poor 
phase coverage, the strong azimuthal field vector contribution
serves as a more accurate constraint in the potential field reconstructions.
As the maps show in Fig.~\ref{fig:simplephase},
we find that this is indeed the case. 
The potential field maps
reconstruct a reduced amount of flux but retain the correct polarity
and show less cross-talk than the ZD maps.
This is also reflected in the $\chi^{2}$ calculations shown in
Table~\ref{tab:chisq1}.

\subsection{Complex images}

\begin{figure*}
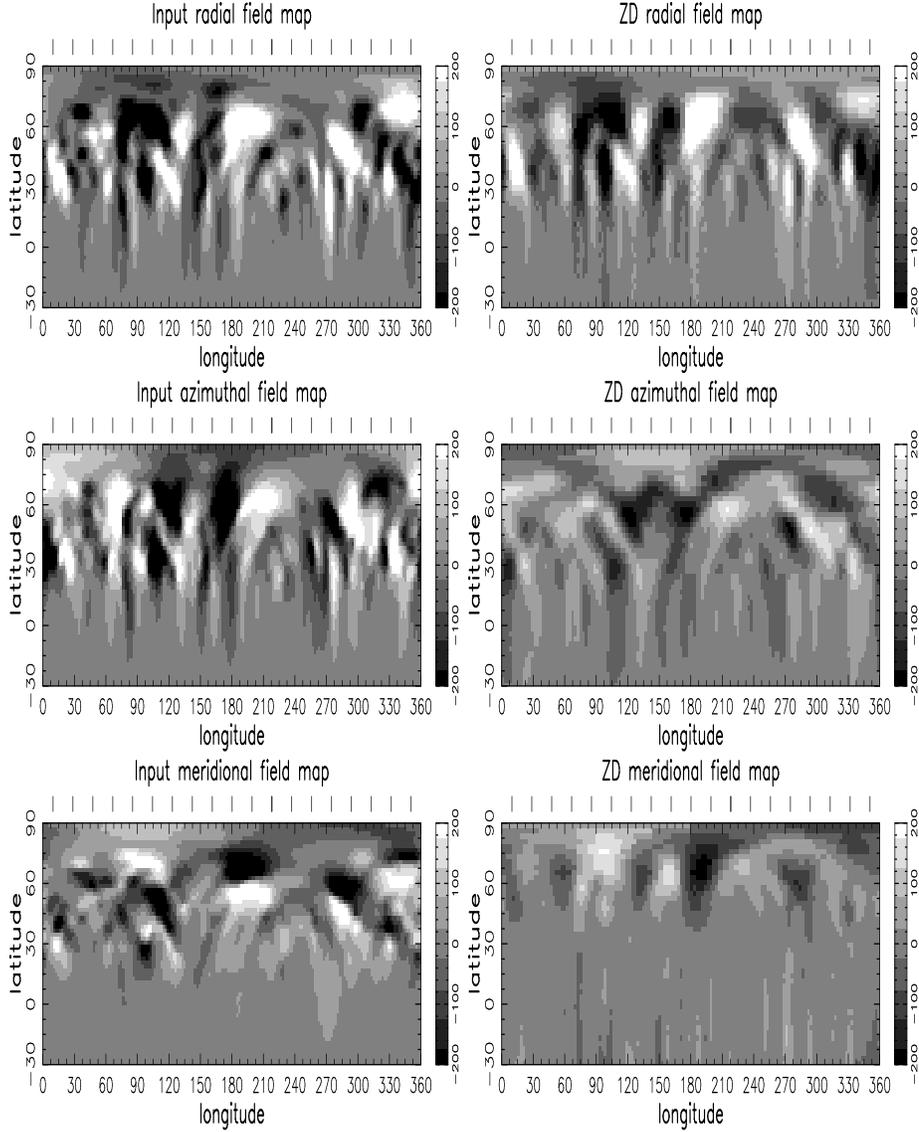


\centerline{\mbox{\psfig{file=fig6a.ps,width=6cm,height=5cm}
 \psfig{file=fig6b.ps,width=6cm,height=5cm}}}
\centerline{\mbox{\psfig{file=fig6c.ps,width=6cm,height=5cm}
 \psfig{file=fig6d.ps,width=6cm,height=5cm}}}
\centerline{\mbox{\psfig{file=fig6e.ps,width=6cm,height=5cm}
 \psfig{file=fig6f.ps,width=6cm,height=5cm}}}
\caption{Input potential surface radial, azimuthal and meridional field maps
calculated using Zeeman Doppler radial field map for AB Dor in 1996 December
(Jardine et al. 1999).
Potential field reconstructions are in the middle column 
and Zeeman Doppler maps are in the right column. 
Greyscales represent the flux in Gauss and 
the tick marks denote the observation phases used to generate model Stokes V spectra as before. }
\label{fig:complexinp}
\end{figure*}

\begin{figure*}
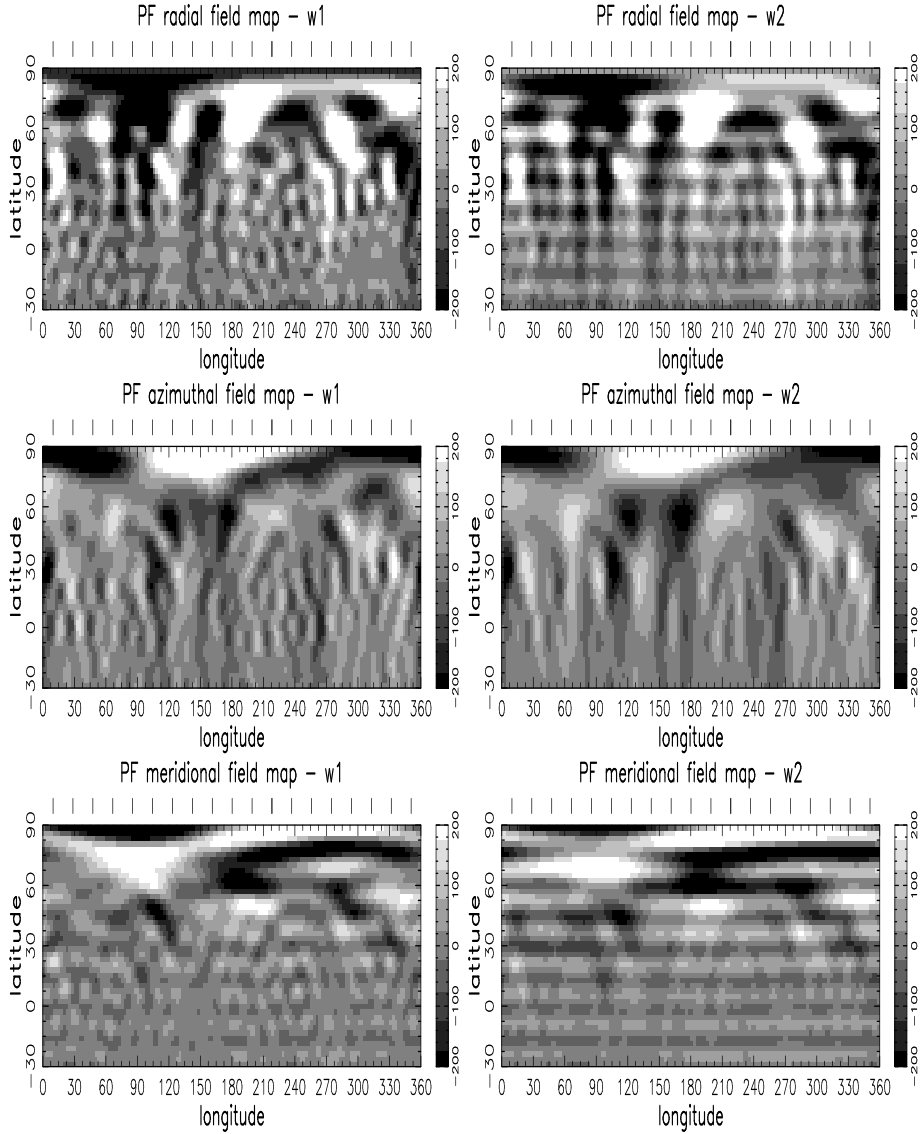


\centerline{\mbox{\psfig{file=fig7a.ps,width=6cm,height=5cm}
\psfig{file=fig7b.ps,width=6cm,height=5cm}}}
\centerline{\mbox{\psfig{file=fig7c.ps,width=6cm,height=5cm}
\psfig{file=fig7d.ps,width=6cm,height=5cm}}}
\centerline{\mbox{\psfig{file=fig7e.ps,width=6cm,height=5cm}
\psfig{file=fig7f.ps,width=6cm,height=5cm}}}
\caption{Reconstructed potential radial, azimuthal and meridional 
field maps using different weighting schemes. The first column shows 
reconstructions using weighting scheme with $n=1$ and the second one
shows maps reconstructed using $n=2$.  Greyscales represent the flux in Gauss 
and the tick marks denote the observation phases used to generate 
model Stokes V spectra as before. }
\label{fig:complexpfs}
\end{figure*}

\begin{figure}
\centerline{\psfig{file=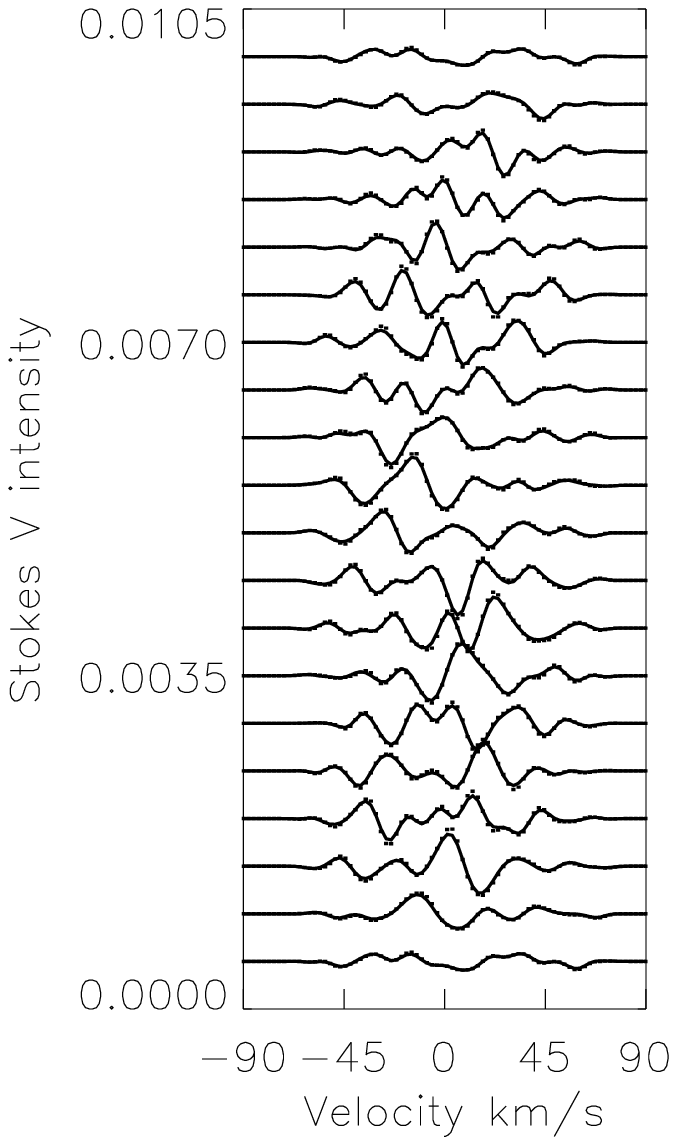}}
\caption{Model input Stokes V spectra generated using the input 
image shown in Fig.~6 are represented by the error bars.
Spectra are plotted in order of increasing phase in steps of 0.05 phase.
The predicted profiles from the potential field reconstructions are
the solid lines. The Zeeman Doppler predicted line profiles
are not plotted here but fit the input spectra to the same degree.}
\label{fig:complexfits}
\end{figure}

We present potential field reconstructions to a more complex,
realistic field distribution in this section.
The input image (Fig.~\ref{fig:complexinp}) 
is of a potential field taken from \scite{jardine99pot}. 
It is the predicted potential field for AB Dor in 1996 December 23-25.
\scite{jardine99pot} produced this image using the ZD 
radial field map from that epoch
and a code originally developed to study the formation of filament
channels on the Sun \cite{vanballegooijen98}.
The model spectra generated using these maps 
are plotted along with the potential field
reconstruction spectral fits in Fig.~\ref{fig:complexfits}.

Fig.~\ref{fig:complexinp} shows that the input and 
reconstructed potential field maps below 
75$^{\circ}$ latitude show good agreement.
At $\theta > 75^{\circ}$, the discrepancy between input and
reconstructed potential field maps increases.
ZD maps also reconstruct more flux near the poles than
is present in the input maps but the disrepancy is not as
great as with the potential field map reconstructions. 
At lower latitudes, on the other hand, it is clear that the 
potential field maps recover more accurate meridional 
field information compared to conventional ZD maps.

\begin{table}
\caption{Reduced chi-squared values evaluating the agreement 
for the reconstructions to the complicated input images.
The first column lists the reconstructed set of images used for the 
calculations: PF-1 stands for weights where $n=1$; PF-2 means
reconstructions where weights $n=2$. 
The rest of the columns indicate the level of
agreement between reconstructed images for each field orientation. 
}

\vspace{0.5cm}
\begin{tabular}{llll}
\hline
Rec. image & $\chi^{2}_{rad}$ & $\chi^{2}_{az}$ & $\chi^{2}_{mer}$ \\
Complex ZD  	& 2.18	& 2.21 &  1.61  \\
Complex PF-2 	& 2.83	& 1.94 &  1.18  \\
Complex PF-1 	& 3.17	& 2.25 &  1.52  \\
\hline
\end{tabular}
\protect\label{tab:chisq2}
\end{table}

It is worth noting that the polar region over which 
potential field maps recover the greatest spurious structure
only represents a small area on the stellar surface.
The addition of this spurious structure 
is most likely due to the definition of the weighting scheme 
(Eqn.~\ref{eq:weights}).
However, the accuracy of the potential field maps in the 
mid to low-latitude regions clearly demonstrates that 
the potential field 
code recovers more information about the surface field 
than is possible using conventional ZDI.
The images produced by different weighting schemes (Fig.~\ref{fig:complexpfs})
show that different weighting schemes redistribute the recovered flux
into different bin sizes on the surface. The overall field pattern is not
changed but does affect the level of fit to the input PF map.
The reduced $\chi^{2}$ measurements for these PF 
reconstructions (Table~\ref{tab:chisq2}) are clearly affected by 
the spurious structure at the pole in the radial field maps. 
However, they do show that in the case of reconstructions using
the area density weighting scheme ($n=2$) the azimuthal and meridional
field reconstructions are more accurate. 

\section{Discussion}

Many limitations in ZDI arise from the sensitivity of
circularly polarized profiles (Stokes V parameter) 
to the line-of-sight components of magnetic fields. 
However, until spectropolarimetric instrumentation is sufficiently
advanced so that all four Stokes parameters can be observed simultaneously
for cool stars this problem is unlikely to be resolved.
Currently, the four Stokes parameters can only be measured in 
bright, highly magnetic A and B stars  \cite{wade00stokes,wade00ap}.

The lack of physical realism in ZD maps also allows for 
unrealistic magnetic field distributions.
By limiting ZD reconstructions to potential field
distributions we can overcome some of these problems.
In mapping potential fields, we introduce 
a mutual dependence on all three surface field vectors thereby
recovering more field information. 
The benefits of this method are threefold:
firstly, in regions where one vector is suppressed 
due to the inclination of the star (such as regions of 
low-latitude meridional field) there is still
a reasonably high contribution from the other two and thus field information
is reconstructed more reliably;
secondly, in areas of bad phase coverage, 
azimuthal fields can sometimes still be recovered (as their signatures 
are strongest about $45^{\circ}$ away from the previous point of observation)
and the mutual dependence of the field vectors 
can be exploited to recover some radial and meridional field information;
and finally, in regions such as starspots where the field is 
thought to be mainly radial but where there is little flux contribution
from all three vectors, the field information in the surrounding
spot areas can be used to recover more of the lost polarized signal.

In addition to these improvements, potential field configurations
should allow us to test if the unidirectional band of azimuthal
field seen in ZD images of AB Dor is necessary to fit the data.
Such a band is incompatible with a potential field and its presence
challenges our current understanding of magnetic field generation.

If datasets are available that span a sufficiently long
time-span it will be possible to monitor which modes are consistently
active on these stars which will give us insight into the 
dynamo mechanisms operating in cool stars.
Extrapolating these surface potential field models to the corona will
enable us to model the coronal topology of the star and thus
to predict where stable prominences may form in these stars.

Finally, special care must be taken when applying this technique to 
rotationally broadened Stokes V spectra from cool dwarfs. 
The weighting scheme used will have to be tested more
thoroughly in order to ensure that we are not sacrificing realism 
for the ``simplest images'' (in terms of information content).
More realistic images may be obtained by incorporating
a form of entropy which pushes the field distribution to a power
law form, a possibility we are investigating further.

\section*{ACKNOWLEDGMENTS}
The image reconstructions were carried out at the
St Andrews node of the PPARC Starlink Project. 
GAJH was funded by PPARC during the course of this work and
MJ acknowledges the support of a PPARC Advanced Fellowship.
We would like to thank Dr A. van Ballegooijen and Prof K.D. Horne
for useful discussions. We are also grateful to the 
referee, Dr S. Saar, for suggestions that have improved the final
version of the paper.

\end{document}

%% file: latex_macros.tex
\newcommand{\rmsub}[2]{#1_{\rm #2}}
%
%
\newcommand{\di}{\mbox{Doppler imaging}}
\newcommand{\zdi}{\mbox{Zeeman-Doppler imaging}}
\newcommand{\sssip}{\mbox{\sc sssip}}
\newcommand{\dotsvn}{\mbox{\sc dots}}
\newcommand{\intense}{\mbox{\sc intense}}
\newcommand{\echomop}{\mbox{\sc echomop}}
\newcommand{\spdecon}{\mbox{\sc spdecon}}
\newcommand{\csev}{\mbox{\em CCP7}}
\newcommand{\aat}{\mbox{\em AAT}}
\newcommand{\memsys}{\mbox{\sc memsys}}
\newcommand{\eso}{\mbox{\em ESO}}
\newcommand{\iue}{\mbox{\em IUE}}
\newcommand{\ucles}{\mbox{\sc ucles}}
\newcommand{\starlink}{\mbox{\em Starlink}}
\newcommand{\etal}{\mbox{\em et\ al.\ }}
\newcommand{\eex}[1]{\hbox{$\hbox{10}^{#1}$}}
\newcommand{\vsini}{\mbox{$v_e\,\sin\,i$}}
\newcommand{\vrad}{\mbox{$v_{\mbox{rad}}$}}
\newcommand{\esprit}{\mbox{\small ESpRIT}}

\newcommand{\ha}{\hbox{$\hbox{H}\alpha$}}
\newcommand{\hb}{\hbox{$\hbox{H}\beta$}}
\newcommand{\hgam}{\hbox{$\hbox{H}\gamma$}}
\newcommand{\heps}{\hbox{$\hbox{H}\epsilon$}}
\newcommand{\lya}{\hbox{$\hbox{Ly}\alpha$}}
\newcommand{\naid}{\mbox{Na~{\sc i} {\sl D}}}
\newcommand{\caii}{\mbox{Ca~{\sc ii}}}
\newcommand{\caiih}{\mbox{Ca~{\sc ii} {\sl H}}}
\newcommand{\caiik}{\mbox{Ca~{\sc ii} {\sl K}}}
\newcommand{\caiihk}{\mbox{Ca~{\sc ii} {\sl H} \&\ {\sl K}}}
\newcommand{\mgii}{\mbox{Mg~{\sc ii}}}
\newcommand{\mgiih}{\mbox{Mg~{\sc ii} {\sl h}}}
\newcommand{\mgiik}{\mbox{Mg~{\sc ii} {\sl k}}}
\newcommand{\mgiihk}{\mbox{Mg~{\sc ii} {\sl h} \&\ {\sl k}}}
\newcommand{\cai}{\mbox{Ca~{\sc i}}}
\newcommand{\fei}{\mbox{Fe~{\sc i}}}
\newcommand{\lii}{\mbox{Li~{\sc i}}}
%
%
\newcommand{\ang}{\,\mbox{\AA}}
\newcommand{\micron}{\,\mbox{$\mu m$}}
\newcommand{\km}{\,\mbox{km}}
\newcommand{\kmsec}{\,\mbox{$\mbox{km}\,\mbox{s}^{-1}$}}
\newcommand{\kev}{\,\mbox{keV}}
\newcommand{\kelvin}{\,\mbox{K}}
\newcommand{\degrees}{\mbox{$^\circ$}}
\newcommand{\rstar}{\,\mbox{$\mbox{R}_*$}}
\newcommand{\mstar}{\,\mbox{$\mbox{M}_*$}}
\newcommand{\lstar}{\,\mbox{$\mbox{L}_*$}}
\newcommand{\vstar}{\,\mbox{$\mbox{V}_*$}}
\newcommand{\msun}{\,\mbox{$\mbox{M}_{\odot}$}}
\newcommand{\rasun}{\,\mbox{$\mbox{R}_{\odot}$}}
\newcommand{\lsun}{\,\mbox{$\mbox{L}_{\odot}$}}
%
%
%
%
\newcommand{\deriv}[2]{\mbox{${{\displaystyle d#1}\over
                       {\displaystyle d#2}}$}} 
\newcommand{\sderiv}[2]{\mbox{${{\displaystyle d^2#1}\over
                       {\displaystyle d#2^2}}$}} 
\newcommand{\pderiv}[2]{\mbox{${{\displaystyle\partial#1}\over
                       {\displaystyle\partial#2}}$}} 
\newcommand{\spderiv}[2]{\mbox{${{\displaystyle\partial^2#1}\over
                        {\displaystyle\partial#2^2}}$}} 
\newcommand{\half}{\mbox{$\frac{1}{2}$}}
\newcommand{\delam}{\mbox{$\Delta\lambda$}}
\newcommand{\bvec}[1]{\mbox{\boldmath ${#1}$}}